\documentstyle[12pt]{article}
\advance \textheight by \topskip
\begin{document}
\hfill{TFP-95-02}
\vspace{0.8cm}
\begin{center}
\begin{large}

        {\bf Andreev tunneling into a one-dimensional
	     Josephson junction array}

\end{large}

\vspace{2cm}

	G. Falci$^{a,c)}$ , Rosario Fazio$^{a,c)}$,
	A. Tagliacozzo$^{b,d)}$ and G. Giaquinta $^{c,d)}$

\vspace{2cm}

	a)Institut f\"ur Theoretische Festk\"orperphysik,
	Universit\"at Karlsruhe, 76128 Karlsruhe, Germany\\
	b) Dipartimento di Scienze Fisiche, Universit\`a di Napoli,
	Mostra d'Oltremare, Pad.19, 80125 Napoli, Italy\\
	c)Istituto di Fisica, Universit\`a di Catania,
	viale A. Doria 6, 95129 Catania, Italy \\
	d)GNSM (CNR) and INFM , Italy\\

\vspace{1cm}

\end{center}

{\bf Abstract}

\vspace{1cm}

In this letter we consider Andreev tunneling between a
normal metal and a one dimensional Josephson junction
array with finite-range Coulomb energy.
The $I-V$ characteristics strongly deviate from the
classical linear Andreev current. We show that the non linear
conductance possesses interesting scaling behavior when the
chain undergoes a $T=0$ superconductor-insulator transition
 of Kosterlitz-Thouless-Berezinskii  type.
When the chain has quasi-long range order, the low lying excitation are
gapless and the $I-V$ curves are power-law (the linear relation is
recovered when charging energy can be disregarded). When the chain is
in the insulating phase the Andreev current is blocked at
a threshold which is
proportional to the inverse correlation length in the chain
(much lower than the Coulomb gap)
and which vanishes at the transition point.

\vspace{2cm}


\newpage

In the past few years there has been a renewed interest in the
properties of meta-superconductor (NS) interfaces~\cite{Beenakker}.
At low temperature and voltage the transport through a
NS interface is due to Andreev tunneling~\cite{Andreev}:
a quasi-electron in the normal metal is
reflected as a quasi-hole and a Cooper pair is added to the
condensate on the superconducting side.
Interference effects have been
predicted for the subgap conductivity in NS tunnel
junctions~\cite{HekkingNazarov}.

Kim and Wen studied the effect of
collective modes below the superconducting gap on the NS
tunneling~\cite{Kim}; due to the excitation of the
low-lying modes in the tunneling process, the $I-V$
characteristics become non-linear.
If the electrostatic energy is large enough,
charging effects suppress transport at low
temperature and voltage: this is the so called Coulomb
blockade (see Ref.\cite{Scharge} for reviews). Tunneling can take place
only if electromagnetic modes with energy of the order of charging
energy can be excited
whose effect on Andreev tunneling has been studied by Hekking et
al.~\cite{Hekkingetal}.

In this letter we will study Andreev tunneling from a normal metal
electrode to a one dimensional Josephson Junction Array (JJA).
The system is depicted in Fig.~1.
The transport in the JJA is due to Josephson tunneling.
The Josephson energy ($\propto E_J$) is minimized for
a phase coherent superconducting state. If charging
effects are relevant, the physics at low
temperatures is dominated by its competition with the electrostatic energy
($E_0=e^2/2C_0$ where $C_0$ is the capacitance of the islands
to the ground). The latter  favours quantum fluctuations of the phase
$\varphi$ of the superconducting order parameter which can destroy
superconductivity~\cite{MooijSchon}.
In one dimensional JJA (which the case we will consider)
this results in a $T=0$
phase transition~\cite{Bradley,Kor-Zwerg-Bob} of the
Kosterlitz-Thouless-Berezinskii type
at $\sqrt{E_J/8E_0} = 2/\pi $.

We will show that
Andreev tunneling into a JJA is a powerful tool to perform
spectroscopy of low lying modes of JJA.
Moreover,
it offers another way to study the critical properties of strongly
interacting superconducting systems. Infact, quantum fluctuations of
the phases in the JJA
produce a nonlinear Andreev current which is very
sensitive to the critical properties of the JJA. Here we point out some
features of the phase diagram of the device which are reflected in
the two-particle tunneling.

Very recently it has been proposed that two electron tunneling
can probe the spectrum of an $S-S$
junction~\cite{HekkingGS}.
The physics of this system is contained also in our
setup in the limit of very small $E_J$. In this limit,
the correlation length
$\xi$ of the JJA will not exceed few lattice constants and the transport
properties are mainly determined by the junctions closest to the
N-S boundary.

We start from the Hamiltonian:
\begin{eqnarray}
&H& =
\sum_{\sigma } \int_{N} d^3r \;\psi ^\dagger _\sigma (\vec{r})\epsilon
(-i\nabla )\psi _\sigma (\vec{r}) +
\frac{1}{2} \sum_{ij} Q_{i}C^{-1}_{ij}Q_{j}
\nonumber\\
&-& \! E_J \sum_i \cos(\varphi_i -\varphi_{i+1})+
t \left[ \mbox{e}^{i\varphi_0 }\psi_\uparrow (0)
\psi _\downarrow(0) +h.c.\right] .
\label{eq:ham}
\end{eqnarray}
It includes the Hamiltonians of the metal particle,
that of the one dimensional JJA and the tunneling Hamiltonian which
describes Andreev processes between the normal metal and the first
superconducting island respectively. In Eq.~(\ref{eq:ham})
$\psi^\dagger  _\sigma $  are the creation operators
for the electrons of the normal metal particle and $\epsilon(-i\nabla
)= -\frac{\nabla ^2}{2m} - \mu $ is
the relative kinetic energy , measured from the Fermi energy. The
integral is extended to the normal metal and $\sigma $ labels the spin index.
$Q_{i}$ and $\varphi_{i}$ denote
the charge and the phase of the i-th island of the JJA and are
canonically conjugated, $[Q_{i},\varphi _{j}]=2e\delta_{ij}i$.
We will consider a model for the
capacitance matrix $C_{ij}$ where
only the ground capacitance $C_{0}$ and the
nearest neighbor capacitance $C$ are present. The range of the
electrostatic interaction between Cooper pairs is
$\lambda =\sqrt{C/C_{0}}$.
In Eq.(\ref{eq:ham}) tunneling occurs with an amplitude $t$ at a single
point of the junction \cite{mutichannel}. Finally we disregard
quasiparticle tunneling in the JJA since we are dealing with
energies far below the superconducting gap.

The Andreev current across the NS boundary when a voltage $V$
is applied is calculated from the time variation of the electron
charge at the metal island
\begin{eqnarray}
I_{NS} &=& -4 e \mid t \mid ^2 \,\, {\cal I} m \, D_R(-2eV)
\nonumber\\
&=& 4 e \mid t \mid ^2 \, \int_{-\infty}^{\infty} dt \;
\mbox{e}^{-2ieVt} \; i \Theta (t)
\; \langle [A^{\dagger }(t),A(0)] \rangle
\label{green}
\end{eqnarray}
where $A(t)=\mbox{e}^{i\varphi _0(t) } \, \psi _\uparrow (0,t)
\psi _\downarrow(0,t)$. The retarded correlator in Eq.~(\ref{green})
can be obtained from the
analytic continuation of the corresponding time
ordered Green function in imaginary frequencies. The latter is the
Fourier transform of
$D_T(\tau )$ $=$ $\langle {\cal T}_{\tau }A^{\dagger }(\tau )A(0) \rangle $
 $=$ $ \langle {\cal T}_{\tau } \, \psi _\uparrow (0,\tau) \, \psi^{\dagger }
 _\uparrow(0,0)\rangle \;
\langle {\cal T}_{\tau } \, \psi _\downarrow (0,\tau) \, \psi ^{\dagger}
 _\downarrow(0,0) \rangle \; g(\tau ) \,$.
All the key features are contained in the phase--phase correlation function
of the first superconducting island,
$
g(\tau ) = \langle {\cal T}_\tau \mbox{e}^{i\varphi_0(\tau )}
\mbox{e}^{-i\varphi_0(0)} \rangle
$.
The critical properties of the JJA are reflected in the form of $g(\tau)$
and can be studied by mapping the ``quantum'' one dimensional JJA at
$T=0$ into a classical $1+1$ dimensional (space and time)
XY-model~\cite{Bradley} where $E_0$ plays the role of the
``temperature''.
Vortices in the classical XY model correspond to $T=0$ phase slips
in the JJA which occur at a certain junction of the chain and are due
to quantum fluctuations modulated by $E_0$.

$\underline{E_0 \gg E_J}$  \hspace{.6cm}
In this limit phase slips destroy the phase order and the Josephson
chain is insulating.
In the disordered phase the correlator $g(\tau)$ decays exponentially in
imaginary time. A
careful evaluation of $g(\tau)$ close to the transition point has been
done in Ref.~\cite{Heinekamp}. At large times, taking
into account the mapping we get
\begin{equation}
g(\tau ) = \mbox{e}^{-(\omega_p \mid \tau \mid )/\xi }/(\omega_p \mid
\tau \mid )^{1/2}  \quad
\tau \to \infty
\label{tinfty}
\end{equation}
where $\omega_p = \sqrt{8 E_J E_0}$. In the limit $C\ll C_0$
the correlation length is given by
$
\xi = \exp [b/\sqrt{ 2/\pi - (E_J/8E_0)^{1/2}}] \,
$, where $b \sim 1$. It
diverges at the superconductor-insulator transition.
Inserting the analytic continuation of Eq.(\ref{tinfty}) into
Eq.(\ref{green}) we get
the tunneling current at small voltages of the array in its
insulating phase
\begin{equation}
I_{NS} = \frac{4G_A \omega _p }{3 e \sqrt \pi } \;
\mbox{sign}(V) \; \Theta \left(
2e \mid V \mid / \omega_p  - \xi  ^{-1} \right)
\; \left[
2e \mid V \mid / \omega_p - \xi ^{-1} \right]^{3/2} \; .
\label{Iins}
\end{equation}
Here $G_A = 8\pi e^2 \mid t \mid ^2 N^2(0)$ is the Andreev
conductance for a
bulk superconductor ($N(0)$ is the density of states
of the metal particle).
The typical $I-V$ characteristics is shown in Fig.~(2)
, curve (a).
The exponential decay of correlation in the JJA corresponds to
the Coulomb blockade of the Andreev tunneling. Interestingly enough,
the threshold is not the bare
charging energy but it is proportional to the inverse correlation length.
Hence it vanishes at the transition with a behavior related
to the critical state of the chain,
$$
V_{tr} = \omega_p/ 2e \xi  \quad .
$$
As long as the correlation length $\xi$ is larger than $\lambda $, the
$I-V$ characteristics are mainly determined by the self-capacitance
and eq.(\ref{Iins}) is valid also for finite $C$.
At lower values of the ratio $E_J/E_0$ further away from
the transition, the
correlation length $\xi $  decreases and eq.(\ref{Iins}) does
not hold any longer because $C$ cannot be disregarded.

At the critical value $\sqrt{E_J/8E_0} = 2/\pi $ the JJA undergoes a
transition to the superconducting phase
where the  phase-slips of opposite sign are tightly bound in
pairs and the voltage drop across the array is zero.

$\underline{E_J \gg E_0}$ \hspace{.6cm}
The Josephson Hamiltonian
is well approximated by its phase-wave Gaussian limit in which
quantum phase slips are neglected.
The low energy excitations are modes with dispersion
\begin{equation}
\omega_q ^2 = \frac{2 \, E_J \, (1-\cos \, q)}
{(8E_0)^{-1} + 2\, (8E_C)^{-1} \,(1-\cos\,q)}.
\label{dispersion}
\end{equation}
The resulting phase correlator in imaginary time is
\begin{equation}
g(\tau )= \exp \left \{ -(\eta -1) \int ^{\infty}_0 \frac{dx}{x}\:
\frac{ 1 - \cos ( x \tilde{\tau})}{\sqrt{1 + x^2 }}\right \}
\label{gtot}
\end{equation}
where $ \tilde{\tau} = \kappa \omega_p \tau$,   $\kappa =
(1/4 + C/C_0)^{-1/2} $ and $\eta = 1+(2 \pi)^{-1}\,\sqrt{8E_0/E_J}$.
The system possesses quasi-long range
order and the correlations decay with a power-law for $\tau \to \infty $,
$
\, g(\tau ) \sim \, \mid \omega_p \tau \mid ^{\,\eta-1} \,
$.
As a result the $I-V$ characteristics at low voltage have a power-law
behavior
\begin{equation}
I_A = G_A \; \frac{ \mbox{e}^{-\gamma (\eta - 1)}}{\Gamma (1 + \eta )} \;\,
\frac{\omega _p}{ 2e}  \;
\left( \frac{ 2e \mid V \mid}{\omega _p} \right)^\eta \quad
V \ll \omega_p/e
\label{swchar}
\end{equation}
where $\gamma $ is the Euler number.
In the classical limit, $\eta \to 1 \,$, the ground state of the chain
has all the phases aligned and no collective modes are excited.
Then the chain behaves as a bulk superconductor and the
Andreev $I-V$ characteristic is linear.
When $E_0$ increases phase-wave collective modes become active.
In particular they are excited by the Andreev tunneling events themselves
and their back-action determines the anomalous nonlinear characteristics
of Eq.~(\ref{swchar}).

The phase-wave Gaussian limit is
the fixed point Hamiltonian also when quantum phase slips are included
but the parameters are renormalized.
In Eq.~(\ref{swchar}) $\eta $ is replaced by the renormalized
value which deviates
from the phase-wave value close to the S-I transition (see
Ref.\cite{Jose} for more details).

Comparing eq.(\ref{Iins}) with eq.(\ref{swchar}) one sees that close to
the transition point the $I-V$ curves start as a power law, but with
different exponents when approaching the transition from the two sides
(see the curves (b) and (c) in Fig.~(2)).
Approaching the transition from above
$\eta _{cr}^+ = 5/4$.
The jump is
$$
\eta _{cr}^- - \eta _{cr}^+ = 1/4
$$
The jump the $I-V$ curves is  related to the jump in
the superfluid density at the transition point.
A similar behavior, though with a different value,
has been predicted and observed long time
ago for the conductance
in the classical 2D superconducting films~\cite{Epstein}.

Now we analize in more detail the Andreev $I-V$ characteristics in the
ordered phase, $\sqrt{E_J/8E_0} > 2/\pi $. Eq.~(\ref{swchar}) is valid at
voltages much smaller than $\omega_p/e$.
For voltages much larger than $ \omega_p /\kappa$ the short time limit of
Eq.(\ref{gtot}) is needed, which is
\begin{equation}
g(\tau )= \exp \left \{ -\frac{\eta -1}{2} \, (\pi \tilde{\tau} -
\tilde{\tau}^2
+  \frac{\pi}{12} \tilde{\tau}^3 ) \right \}
\quad
\tau \to 0
\label{gshort}
\end{equation}
Analytic continuation leads to
\begin{equation}
I_{NS} = G_A \left( V - \frac{\pi}{2} \, (\eta-1) \,
\frac{\kappa \omega_p }{e}
\right)
 \hspace{1cm} V \gg \frac{\kappa \omega_p }{e}
\label{ilarge}
\end{equation}
with corrections which vanish exponentially with the applied voltage.
At very large voltages, where eq.(\ref{ilarge}) applies,
we recover the linear behavior of the $I-V$ characteristics, apart
from an offset which is related to the average energy cost
for exciting the collective phase-wave modes of the chain. The $I-V$
curves are shown in the inset of Fig.~(2). We calculated the
behavior at intermediate voltages directly in real times from
Eq.(\ref{green}).
The excitations of the chain act as bosonic modes of an electromagnetic
environment \cite{Ingold} coupled to tunneling pair and their energy
spectrum is given by Eq~(\ref{dispersion}). Integrating the Bose operators
of the environment we get the real time correlator
\begin{equation}
P(t) = \langle e^{i\varphi_0(t )}
e^{-i\varphi_0(0)} \rangle =
 \exp \left\{ - \int_{-\pi }^{\pi } \frac{dq }{4\pi } \;
 \zeta (q) \;
\left[ 1 - \cos(\omega_q t) + i \sin(\omega_q t) \right] \right\}
\label{Pomega}
\end{equation}
where
$\zeta (q) = \sqrt{e^2C^{-1}(q)/E_J\sin^2(q/2)}$.
The current takes the form
\begin{equation}
I_{NS} = -4 e \mid t \mid ^2 N^2(0) \int_{0}^{2eV} \frac{d\omega }{2\pi }
\;	(2eV - \omega ) \; P(\omega ) \; .
\label{isup}
\end{equation}
The power law behavior of Eq.~(\ref{swchar}) for small voltages
turns very rapidly
in the asymptotic form, Eq.~(\ref{ilarge}), at
$V \approx \kappa \omega_p /e$.
$P(\omega)$ can be interpreted as the probability that a tunneling event
excites a mode of the JJA. This results resembles that obtained in
connection with the study of the effect of the electromagnetic
environment on single electron
tunneling~\cite{Ingold}.
In this approximation the one-dimensional JJA acts like a line of linear
elements with the equivalent impedance
\begin{equation}
\frac{Z(\omega)}{h/2e^2} = \frac{\eta-1}
	{\sqrt{1-(\omega/\kappa \omega_p)^2}} \;
	\Theta \left(\kappa \omega_p - \omega \right) \; .
\label{Zomega}
\end{equation}
In the phase-wave approximation the function $P(\omega)$ presents
an elastic peak, $P(\omega) \sim \omega^{\eta-2}$ for $\omega \to 0$,
and a divergence at $\omega = \kappa \omega_p$,
$P(\omega) \sim |\kappa \omega_p - \omega|^{\eta-3/2}$. This latter
feature is associated with excitations which are almost localized in
some junction of the JJA and whose density of states diverges for the
infinite chain. It produces some structure near
$V = \kappa \omega_p/e$ both in $\partial ^2
I / \partial ^2 V \sim P(eV)$ and in the
differential conductance $\partial I / \partial V$ which
is shown in the inset of Fig.~(3).
Notice that the low voltage behaviour is dominated by the
capacitance to the ground $C_0$ since $C$ enters simply in the scaling
factor $\kappa$ (see Eqs.(\ref{ilarge},\ref{Zomega})).

The modifications we expect in going beyond the phase-wave approximations
in the ordered state can be qualitatively discussed in terms of the
spectrum of the excitations we neglected. Creation and breathing of
bound pairs of phase slips produce low energy Gaussian fluctuations
which, as already discussed, only renormalize $\eta $ and modify
the exponent in Eq.~(\ref{swchar}) accordingly. They should not affect
the structure near $V = \kappa \omega_p/e$ which is due to localized
excitations. The latter, however, can be modified due to the
interaction of these modes with the phase slips or between themselves.
Moreover going to higher voltages phase slips may be induced. Their
typical energy scale, away from the
transition, is larger than $\kappa \omega_p$ so in this
regime {\it source induced} phase
slips are believed not to be important. They should start to affect
the behavior at $V \sim \kappa \omega_p/e$ only close to the
transition, so we expect a detectable region in the phase diagram
where the behavior of Eq.~(\ref{isup}) can be observed.
At even higher voltage all the above excitations can be active and
this should increase the offset in the $I-V$ curve, Eq.~(\ref{ilarge}).

In conclusion we have shown that Andreev tunneling may be a
powerful method to investigate the
phase diagram and the excitation spectrum (see also \cite{HekkingGS})
of JJA.
The results derived in this letter are summarized in
Figs.~(2,~3).
When the JJA is in the disordered phase the Andreev tunneling is
blocked up to a threshold voltage proportional
to the correlation length $\xi $ (Fig.~(2)).
At the superconductor-insulator transition the $I-V$ curves are
power law like with a jump in the exponent at the transtion
point (Fig.~(2)). When the JJA is in the superconducting phase,
the characteristics start as a power law and approach rapidly the shifted
ohmic behaviour. A richer structure generated by localized
excitations of the JJA is seen in the derivatives of the
$I-V$ curves (see Fig.~(3)),
near $V = \kappa \omega_p /e$.
We have studied explicitely a one dimensional JJA but other types of arrays
can be treated along the same lines.
In two dimensions frustration effects due to an
applied magnetic field play an important role. The
interference patterns manifest in the differential conductance
in a rather peculiar way. Work in this direction is in progress.
\\
{\bf Acknowledgements} We thank C. Bruder and G. Sch\"on for useful comments.
We acknowledge financial support from
European Community under contract HCM-network CHRX-CT93-0136
and (for G.F.) under EC grant ERBCHBICT930561.

\newpage

\noindent
{\bf FIGURE CAPTIONS}

\vspace{2cm}

\noindent
Figure 1:The system considered in the present work: a normal metal
island connected by a tunnel junction to a very long chain of Josephson
junctions.

\vspace{0.7cm}
\noindent
Figure 2:(a) The $I-V$ curve in the limit $E_0 \gg E_J $: $I=0$ for
$V<V_{th}$. At the transition point the current starts as a power law,
however there is a jump in the exponent of the characteristics
(curves (b) and (c)). For $E_J \to \infty$ the classical linear behavior
is recovered (d). In the inset some curves in the $E_J \gg E_0$
regime.

\vspace{0.7cm}
\noindent
Figure 3:The function $P(\omega)$, which also gives the second
derivative of the  $I-V$ curves, in the limit $E_J \gg E_0$:
$E_0/E_J = 0.05,0.02,0.005,0.0005$, from top to bottom.
In the inset the differential conductance, in the same limit,
is suppressed for increasing $E_0/E_J$.


\begin{thebibliography}{99}
\bibitem{mutichannel}	The results presented in this paper can be
	applied also to fabricated junctions, in wich tunneling occurs
	via many independent channels.
\end{thebibliography}
\end{document}